# ESTIMATION OF GENETIC RISK FUNCTION WITH COVARIATES IN THE PRESENCE OF MISSING GENOTYPES

ANNIE J. LEE$^A$, KAREN MARDER$^{B,C}$, ROY N.ALCALAY$^{B,C}$, HELEN MEJIA–SANTANA$^B$, AVI ORR–URTREGER$^{D,E}$, NIR GILADI$^{F,G}$, SUSAN BRESSMAN$^H$, YUANJIA WANG$^G$

ABSTRACT. In genetic epidemiological studies, family history data are collected on relatives of study participants and used to estimate the age-specific risk of disease for individuals who carry a causal mutation. However, a family member's genotype data may not be collected due to the high cost of in-person interview to obtain blood sample or death of a relative. Previously, efficient non-parametric genotype-specific risk estimation in censored mixture data has been proposed without considering covariates. With multiple predictive risk factors available, risk estimation requires a multivariate model to account for additional covariates that may affect disease risk simultaneously. Therefore, it is important to consider the role of covariates in the genotype-specific distribution estimation using family history data. We propose an estimation method that permits more precise risk prediction by controlling for individual characteristics and incorporating interaction effects with missing genotypes in relatives, and thus gene-gene interactions and gene-environment interactions can be handled within the framework of a single model. We examine performance of the proposed methods by simulations and apply them to estimate the age-specific cumulative risk of Parkinson's disease (PD) in carriers of *LRRK2* G2019S mutation using first-degree relatives who are at genetic risk for PD. The utility of estimated carrier risk is demonstrated through designing a future clinical trial under various assumptions. Such sample size estimation is seen in the Huntington's disease literature using the length of abnormal expansion of a CAG repeat in the *HTT* gene, but is less common in the PD literature.

## 1. Introduction

Estimating the cumulative risk of disease onset by a certain age for individuals who carry a causal mutation (i.e., age-specific penetrance function), has important implications for both genetic counseling and clinical trial planning. For example, the age-specific penetrance function will provide prognostic information for patients who are at risk of a genetic disorder wishing to know the significance of their biological features in relation to their risk of disease-onset. This will ultimately allow genetic counselors to interpret risk of disease according to each individual's unique biological characteristics and help patients make important decisions regarding genetic testing. Furthermore, cumulative disease risk obtained from an untreated population provides an estimate of the baseline event rate when powering a clinical trial recruiting individuals at genetic risk. A lack of treatment options for changing the trajectory of Parkinson's disease (PD) or Alzheimer's disease (AD) progression, in combination with an increasing elderly population, poses a rising economic burden on patients and the healthcare system [1], which makes developing innovative new treatments that delay disease onset an urgent research priority [1, 2]. In many genetic epidemiological studies of these late-onset disorders, family history data on probands at genetic risk are collected and used to estimate the penetrance [3, 4, 5, 6]. In a typical study, affected individuals (e.g., Ashkenazi Jewish (AJ) PD patients) including carriers and non-carriers are recruited, and report family history of disease, including age at onset of the disease, in their first-degree relatives. However, genotype information in many relatives may not be available due to death of a relative (e.g., parents of a proband) or lack of resources to collect blood samples in all family members [5, 7, 8]. This practical difficulty is frequently encountered in cases of late onset disease (e.g., PD or AD) when parents are often deceased. When some genotype information is missing, the probability of a family member carrying a mutation is estimated based on the mutation status in the initial cohort of subjects (e.g., probands) and the relative's relationship to the proband under Mendelian transmission. [3, 5, 8].

With missing genotypes and ages of disease onset subject to right censoring, the observed data consists of censored mixture data. Previously, a sieve maximum likelihood nonparametric method for such data was proposed to estimate distribution function in carriers and non-carriers [9]. The method was applied to a recent study [5] to estimate age-specific risk of PD in *LRRK2* gene mutation carriers compared to non-carriers. To evaluate the sex effect on the carrier risk in this study, stratified analyses were carried out and the penetrance function was estimated in a subpopulation of male and female relatives separately [5]. Stratified analysis maintains the nonparametric nature in the sense that it does not assume any specific model of the association between the covariates and the outcome. However, a limitation is reduced efficiency; when multiple covariates are available that may affect the penetrance simultaneously, stratifying by a larger number of covariates reduces the sample size and estimation may become infeasible. Therefore, when considering the role of other demographic covariates or environmental risk factors on modifying the penetrance functions, it is desirable to link covariates to the penetrance function through appropriate semiparametric regression models.

Another advantage of introducing covariates to the model is that gene by gene interaction or gene by environmental risk factor interaction can be handled within the framework of a single model. A better characterization of the interactions between genetic and environmental factors helps to understand the pathogenesis of multifactorial diseases [10]. However, if the gene itself is examined without considering its potential interactions with other factors, the effect of the genetic factors on complex mechanism might be missed [11]. For example for PD, variants in *LRRK2* and *PARK16* genetically interact to increase the risk of PD. A recent study shows that *LRRK2* interacts with *RAB7L1* to modify PD risk [12, 13]. Therefore, it is of interest to consider genotype-specific risk estimation in the presence of their interaction with other genetic or environmental factors. A *LRRK2* by gender interaction was suggested in a prior stratified analysis in our motivating study [5]. It is desirable to test for this interaction in a parsimonious model. One challenge in testing for gene by other risk factor interaction in [5] and other similar studies [4] is that genotypes are not available in most first-degree relatives.

In this paper, we propose a covariate-adjusted semi-parametric estimation method that permits including multiple covariates and interaction effects in the presence of missing genotypes through a semiparametric regression model. Compared to previous nonparametric approaches, our method allows controlling for individual characteristics such as sex, ethnicity, environmental risk factors, and genotypes at other loci. Moreover, gene-gene interactions and gene-environment interactions can also be handled within the framework of a semiparametric model. Thus, we extend the prior work [9] on a single gene to handle multiple genes. The analyses may provide insights on whether demographics or environmental variables play a role in modifying the penetrance. In addition, to assist with clinical trial planning, we estimate cumulative risk distribution of disease onset in the overall sample by marginalizing over covariate distributions using estimates obtained from the conditional model given covariates. Sample size estimation has been calculated in the Huntington's disease literature due to the known disease causal gene and its near complete the penetrance. However, using estimated mutation risk to design a clinical trial for PD is less common due to reduced penetrance. Our estimate of *LRRK2* penetrance in the AJ population provides a unique opportunity to power a future clinical trial in this population with a higher PD risk. When external or prior information on the covariate distribution in the target trial population is available, they can be easily incorporated. The performance of the proposed method is examined through extensive simulation studies. Finally, we apply the proposed approach to estimate the age-specific risk of PD for first-degree relatives with *LRRK2* mutations [4, 14, 15] and test for interaction between *LRRK2* mutation and other covariates.

## 2. Methods

2.1. **Notation and likelihood function.** Let $T_i$ be the age-at-onset of a disease which is subject to random censoring given covariates. Let $X_i$ indicate the carrier status at the causal gene of interest, with one indicating the carrier group (each individual has at least one copy of the mutation) and zero indicating the non-carrier group under an autosomal dominant inheritance model. Note that *LRRK2* G2019s mutation has an autosomal dominant mode of inheritance [5]. Thus, here we rely on biological knowledge and in the following consider a dominant model. It is straightforward to extend to other genetic models (recessive or additive). Let $\boldsymbol{Z}_i$ and $\boldsymbol{W}_i$ be a vector of auxiliary covariates collected on the probands and relatives, respectively. For example, $\boldsymbol{Z}_i$ may include demographic information on probands in each family representing family-specific covariates to be adjusted, and $\boldsymbol{W}_i$ may include relatives' own individual-specific characteristics. Due to potential right-censoring of the age-at-onset information and unknown $X_i$, the observed data from $n$ subjects consist of $\{Y_i = T_i \bigwedge C_i, \Delta_i = I(T_i \leqslant C_i), \boldsymbol{Z}_i, \boldsymbol{W}_i\}, i = 1, \ldots, n$, where $C_i$ denotes the censoring time assumed to be conditionally independent of $T_i$ given $\boldsymbol{Z}_i$ and $\boldsymbol{W}_i$.

In our motivating study, the sampling design was to first recruit probands (either affected by the disease or control) in a three-site PD consortium studying *LRRK2* G2019S [5, 9]. All the probands are genotyped and their phenotype information and covariates were collected ($\boldsymbol{Z}_i$ and $Y_i$ are known). Next, the family history of the disease and some demographic information of the first-degree relatives, were collected through a systematic interview with the probands or the relatives themselves [7] ($T_i$ and $\boldsymbol{W}_i$ are known). Genotypes on most relatives were not available ($X_i$ can be missing). Even though in some cases the relatives' genotypes are unknown, one can estimate the probability of a relative being a carrier, i.e., $P(X_i = 1)$. For example, *LRRK2* gene mutation is associated with autosomal dominant PD [16] and knowing just one parent is homozygous carrier will lead to infer the child is a carrier with probability one. Moreover, a child of a heterozygote carrier parent has a probability of 0.5 of carrying this mutation under the Mendelian transmission and a low mutation prevalence (approximately zero). If the mutation prevalence in the general population is known and denoted by $c$, then the probability of this child being a carrier is $P(X_i = 1) = 0.5(1+c)$. To present the likelihood function, we denote a finite number of possible values for $P(X_i)$ in a study as $\{p_1, \ldots, p_m\}$, and let an indicator variable, $G_i$, represent $m$ distinct carrier probabilities, where $G_i = g$ indicates $P(X_i = 1) = p_g, g = 1, \ldots, m$. The observed genotypes in relatives can be incorporated by letting $p_g$ equal one for carriers and zero for non-carriers. The model identifiability conditions on $p_g$ were examined in Lemma 1 of [9].

Because the mutation status $X_i$ may be unknown, the observed conditional likelihood takes a mixture form as

$$\prod_{i=1}^{n} \prod_{g=1}^{m} \left\{ [p_g f_1(Y_i; \boldsymbol{Z}_i, \boldsymbol{W}_i) + (1-p_g) f_0(Y_i; \boldsymbol{Z}_i, \boldsymbol{W}_i)]^{\Delta_i} \right.$$
$$\left. \times [1 - p_g F_1(Y_i; \boldsymbol{Z}_i, \boldsymbol{W}_i) - (1-p_g) F_0(Y_i; \boldsymbol{Z}_i, \boldsymbol{W}_i))]^{1-\Delta_i} \right\}^{I(G_i = g)},$$

where $f_k(y; \boldsymbol{z}, \boldsymbol{w})$ is the conditional probability density function of $T$ in the group with $X = k$ given covariates $\boldsymbol{Z} = \boldsymbol{z}$ and $\boldsymbol{W} = \boldsymbol{w}$, and $F_k(y; \boldsymbol{z}, \boldsymbol{w})$ is the corresponding cumulative distribution function. Compared to the prior work [9], the main contribution here is to accommodate covariates, gene by environmental risk factor interaction, and potential gene by gene interaction, through regression models. Thus, consider the hazard model allowing for the interaction between $X_i$ and $\boldsymbol{W}_i$ as

(1) $$\lambda(t|X_i, \boldsymbol{W}_i, \boldsymbol{Z}_i) = \lambda_0(t) \exp\{\beta(t)X_i + \boldsymbol{\eta}^T \boldsymbol{W}_i + \boldsymbol{\theta}^T \boldsymbol{W}_i X_i + \boldsymbol{\gamma}^T \boldsymbol{Z}_i\}.$$

Note that we allow the hazard ratio between two genotype groups to be time-dependent (i.e., $\beta(t)$ is time-varying) to accommodate the potential time-varying mutation effect and protect against misspecification of the penetrance function. To adjust for $\boldsymbol{W}_i$, $\boldsymbol{Z}_i$ and interaction effect in a parsimonious way, their effects are assumed to be time-invariant. It is possible to extend to time-varying case. The observed data likelihood is

$$\prod_{i=1}^{n} \prod_{g=1}^{m} \left\{ \left[ p_g \lambda_0(Y_i) e^{\beta(Y_i) + (\boldsymbol{\eta}+\boldsymbol{\theta})^T \boldsymbol{W}_i + \boldsymbol{\gamma}^T \boldsymbol{Z}_i} \exp\left\{ -\int_0^{Y_i} e^{\beta(t) + (\boldsymbol{\eta}+\boldsymbol{\theta})^T \boldsymbol{W}_i + \boldsymbol{\gamma}^T \boldsymbol{Z}_i} d\Lambda_0(t) \right\} \right. \right.$$
$$\left. + (1 - p_g) \lambda_0(Y_i) e^{\boldsymbol{\eta}^T \boldsymbol{W}_i + \boldsymbol{\gamma}^T \boldsymbol{Z}_i} \exp\{-\Lambda_0(Y_i) e^{\boldsymbol{\eta}^T \boldsymbol{W}_i + \boldsymbol{\gamma}^T \boldsymbol{Z}_i}\} \right]^{\Delta_i}$$
$$\times \left[ p_g \exp\left\{ -\int_0^{Y_i} e^{\beta(t) + (\boldsymbol{\eta}+\boldsymbol{\theta})^T \boldsymbol{W}_i + \boldsymbol{\gamma}^T \boldsymbol{Z}_i} d\Lambda_0(t) \right\} \right.$$
(2) $$\left. \left. + (1 - p_g) \exp\{-\Lambda_0(Y_i) e^{\boldsymbol{\eta}^T \boldsymbol{W}_i + \boldsymbol{\gamma}^T \boldsymbol{Z}_i}\} \right]^{1-\Delta_i} \right\}^{I(G_i = g)}.$$

We can further expand the model to accommodate penetrance estimation for more than one gene. For example, in our motivating *LRRK2* consortium study, probands were genotyped for the *LRRK2* G2019S and glucocerebrosidase (*GBA*) mutations, but the probands who carried *GBA* mutation were excluded [5] in prior analyses to investigate the *LRRK2* G2019S mutation effect on the PD-onset that is not due to the other genetic factors. We can now further estimate the risk of PD accounting for multiple genotypes such as cumulative risk for those carrying both *LRRK2* and *GBA* mutation. Let $U_i$ denote the potentially unobserved *GBA* carrier status. The hazard function corresponding to two genes is expanded as

$$\lambda(t|X_i, U_i, \boldsymbol{W}_i, \boldsymbol{Z}_i) = \lambda_0(t) \exp\{\beta_1(t) X_i + \beta_2(t) U_i + \boldsymbol{\eta}^T \boldsymbol{W}_i + \boldsymbol{\theta}^T \boldsymbol{W}_i X_i + \boldsymbol{\gamma}^T \boldsymbol{Z}_i\},$$

where $\beta_j(t)$ is the effect of each gene on the disease hazard, respectively. An interaction term between genes $X_i$ and $U_i$ can also be incorporated when of interest. Because the majority of the relatives were neither genotyped for *LRRK2* nor for *GBA*, we expand the probability vector of the carrier status at two loci as $P(X_i = g_1, U_i = g_2) = p_{g_1, g_2}$. The likelihood function in (2) can be re-expressed under the expanded hazard model and probability vectors, and the estimation procedure proceeds as in a single gene case.

2.2. **Sieve maximum likelihood estimation with covariates and interactions.** To estimate the parameters $\Lambda_0(t)$ and $\{\beta(t), \boldsymbol{\eta}, \boldsymbol{\theta}, \boldsymbol{\gamma}\}$, we adapt a hybrid approach involving a nonparametric estimator and sieve estimation that leads to consistent and semiparametrically efficient estimators similar to [9]. Specifically, consider using a nonparametric maximum likelihood estimator for $\Lambda_0(t)$ and a sieve approximation to estimate $\beta(t)$ by letting $\beta(t) = \sum_{j=1}^{K_n} \alpha_j \phi_j(t)$, where $\phi_1(t), \ldots, \phi_{K_n}(t)$ are basis functions such as B-splines functions. It is computationally intensive and inefficient to directly maximize (2) over all the parameters, since the log-likelihood is not convex and the parameters include the potentially large number of jumps of $\Lambda_0$. However, by treating the mutation status in all individuals $X_1, \ldots, X_n$ as missing data, fast numerical computation can be achieved by using the expectation-maximization (EM) algorithm [17] due to available closed-form solutions in the M-step.

The complete data log-likelihood function for $(Y_i, \Delta_i, X_i, G_i), i = 1, ..., n$ is given as

$$\sum_{i=1}^{n} I(X_i = 1) \left\{ \Delta_i \log \delta \Lambda_0(Y_i) + \Delta_i \sum_{j=1}^{k_n} \alpha_j \phi_j(Y_i) + \Delta_i (\boldsymbol{\eta} + \boldsymbol{\theta})^T \boldsymbol{W}_i + \Delta_i \boldsymbol{\gamma}^T \boldsymbol{Z}_i \right. $$

$$\left. - \sum_{Y_k \leq Y_i} \delta \Lambda_0(Y_k) \exp\{\sum_{j=1}^{k_n} \alpha_j \phi_j(Y_k) + (\boldsymbol{\eta} + \boldsymbol{\theta})^T \boldsymbol{W}_i + \boldsymbol{\gamma}^T \boldsymbol{Z}_i\} \right\}$$

$$+ \sum_{i=1}^{n} I(X_i = 0) \left\{ \Delta_i \log \delta \Lambda_0(Y_i) + \Delta_i \boldsymbol{\eta}^T \boldsymbol{W}_i + \Delta_i \boldsymbol{\gamma}^T \boldsymbol{Z}_i - \Lambda_0(Y_i) \exp\{\boldsymbol{\eta}^T \boldsymbol{W}_i + \boldsymbol{\gamma}^T \boldsymbol{Z}_i\} \right\}$$

$$+ \sum_{i=1}^{n} \sum_{g=1}^{m} I(G_i = g, X_i = 1) \log p_g + \sum_{i=1}^{n} \sum_{g=1}^{m} I(G_i = g, X_i = 0) \log(1 - p_g).$$

Therefore, in the E-step of EM algorithm, we evaluate the posterior probability of $X_i = 1$ given the observation data $(G_i, Y_i, \Delta_i)$ as $q_i = a_i/(a_i + b_i)$, where

$$a_i = p_{G_i} \exp\{\Delta_i \sum_{j=1}^{k_n} \alpha_j \phi_j(Y_i) + \Delta_i (\boldsymbol{\eta} + \boldsymbol{\theta})^T \boldsymbol{W}_i + \Delta_i \boldsymbol{\gamma}^T \boldsymbol{Z}_i - \int_0^{Y_i} e^{\sum_{j=1}^{k_n} \alpha_j \phi_j(t) + (\boldsymbol{\eta} + \boldsymbol{\theta})^T \boldsymbol{W}_i + \boldsymbol{\gamma}^T \boldsymbol{Z}_i} d\Lambda_0(t)\}$$

$$b_i = (1 - p_{G_i}) \exp\{\Delta_i \boldsymbol{\eta}^T \boldsymbol{W}_i + \Delta_i \boldsymbol{\gamma}^T \boldsymbol{Z}_i - \Lambda_0(Y_i) \exp\{\boldsymbol{\eta}^T \boldsymbol{W}_i + \boldsymbol{\gamma}^T \boldsymbol{Z}_i\}\}.$$

In the M-step, we maximize

$$\sum_{i=1}^{n} q_i \left\{ \Delta_i \log \delta \Lambda_0(Y_i) + \Delta_i \sum_{j=1}^{k_n} \alpha_j \phi_j(Y_i) + \Delta_i (\boldsymbol{\eta} + \boldsymbol{\theta})^T \boldsymbol{W}_i + \Delta_i \boldsymbol{\gamma}^T \boldsymbol{Z}_i \right.$$

$$\left. - \sum_{Y_k \leq Y_i} \delta \Lambda_0(Y_k) \exp\{\sum_{j=1}^{k_n} \alpha_j \phi_j(Y_k) + (\boldsymbol{\eta} + \boldsymbol{\theta})^T \boldsymbol{W}_i + \boldsymbol{\gamma}^T \boldsymbol{Z}_i\} \right\}$$

$$(3) \quad + \sum_{i=1}^{n} (1 - q_i) \left\{ \Delta_i \log \delta \Lambda_0(Y_i) + \Delta_i \boldsymbol{\eta}^T \boldsymbol{W}_i + \Delta_i \boldsymbol{\gamma}^T \boldsymbol{Z}_i - \Lambda_0(Y_i) \exp\{\boldsymbol{\eta}^T \boldsymbol{W}_i + \boldsymbol{\gamma}^T \boldsymbol{Z}_i\} \right\}.$$

By differentiating (3) with respect to the jump sizes of $\Lambda_0(\cdot)$ at the observed event times, we obtain close-form solutions

$$\delta \Lambda_0(Y_i) = \frac{\Delta_i}{\sum_{k=1}^{n} I(Y_k \geq Y_i) \left\{ q_k \exp\{\sum_{j=1}^{k_n} \alpha_j \phi_j(Y_i) + (\boldsymbol{\eta} + \boldsymbol{\theta})^T \boldsymbol{W}_k + \boldsymbol{\gamma}^T \boldsymbol{Z}_k\} + (1 - q_k) \exp\{\boldsymbol{\eta}^T \boldsymbol{W}_k + \boldsymbol{\gamma}^T \boldsymbol{Z}_k\} \right\}}. \quad (4)$$

After substituting (4) into (3) and differentiating with respect to $\alpha_j$, we obtain the score equation for $\alpha_j$ as

$$\sum_{i=1}^{n} \Delta_i \left[ q_i - \frac{\sum_{k=1}^{n} I(Y_k \geq Y_i) q_k \exp\{\sum_{j=1}^{k_n} \alpha_j \phi_j(Y_i) + (\boldsymbol{\eta} + \boldsymbol{\theta})^T \boldsymbol{W}_k + \boldsymbol{\gamma}^T \boldsymbol{Z}_k\}}{\sum_{k=1}^{n} I(Y_k \geq Y_i) \left\{ q_k e^{\sum_{j=1}^{k_n} \alpha_j \phi_j(Y_i) + (\boldsymbol{\eta} + \boldsymbol{\theta})^T \boldsymbol{W}_k + \boldsymbol{\gamma}^T \boldsymbol{Z}_k} + (1 - q_k) e^{\boldsymbol{\eta}^T \boldsymbol{W}_k + \boldsymbol{\gamma}^T \boldsymbol{Z}_k} \right\}} \right] \boldsymbol{\phi}(Y_i) = 0,$$

where $\boldsymbol{\phi}(Y_i) = (\phi_1(Y_i), \cdots, \phi_{k_n}(Y_i))^T$. The score equations for the other parameters in the M-step are obtained similarly and the parameters are estimated using Newton-Raphson algorithm. After obtaining the updated $\alpha$ and $\{\boldsymbol{\eta}, \boldsymbol{\theta}, \boldsymbol{\gamma}\}$ values, we use (4) to update the jumps of $\Lambda_0(\cdot)$. We then iterate between the E-step and M-step till convergence. Note that the baseline hazard function has a closed-form solution due to the use of nonparametric maximum likelihood estimator, the M-step is computationally fast. Asymptotic properties of the baseline cumulative hazard function in a nonparametric model were studied in Wang et al. (2015) [9], which shows consistency, efficiency, and convergence to a Gaussian process in the presence of missing genotypes. Similar arguments (see for example, [18]) can be applied to establish asymptotic properties of estimators considered here.

To inform planning of a clinical trial, the marginal cumulative risks in carrier and non-carrier relatives may provide design parameters (e.g., risk of disease in the absence of intervention). Thus, we can estimate the marginal cumulative distribution of disease age-at-onset in carriers and non-carriers, respectively, based on the expressions

$$F_1(t) = P(T \leqslant t | X = 1) = 1 - \int \exp\{-\Lambda_0(t) e^{\beta(t) + (\boldsymbol{\eta} + \boldsymbol{\theta})^T \boldsymbol{W} + \boldsymbol{\gamma}^T \boldsymbol{Z}}\} dH(\boldsymbol{W}, \boldsymbol{Z} | X = 1),$$

and

(5) $$F_0(t) = P(T \leqslant t | X = 0) = 1 - \int \exp\{-\Lambda_0(t) e^{\boldsymbol{\eta}^T \boldsymbol{W} + \boldsymbol{\gamma}^T \boldsymbol{Z}}\} dH(\boldsymbol{W}, \boldsymbol{Z} | X = 0),$$

where $H(\boldsymbol{W}, \boldsymbol{Z} | X)$ denotes the conditional probability distribution function of $\boldsymbol{W}, \boldsymbol{Z}$ given the mutation carrier status. Since the genotypes $X$ are not observed in some relatives, we estimate $H(\boldsymbol{W}, \boldsymbol{Z} | X)$ through an EM algorithm.

## 3. SIMULATION STUDIES

We conducted simulation studies to assess the performance of the proposed method. The simulations were designed to imitate the *LRRK2* penetrance study described in Section 4. We generated survival times similar to the estimated distributions of the actual data in Section 4. The distribution of the event times for non-carriers in the baseline group (with covariates $\boldsymbol{Z}_i$ and $\boldsymbol{W}_i$ as zero) were generated from $Weibull(5, 105)$ and the distributions for carriers and non-carriers in the other covariate groups were generated under the hazard model (1), allowing for relative's sex by gene interaction [9]. The mutation probability $p_i$ was taken from $\{0, 0.02, 0.51, 1\}$, as in the real data analysis. 93% of the total sample size $n = 2266$ were not genotyped and parents, siblings, and children had a similar rate of available genotypes as in the real example. To evaluate how well our method works in the case of data sets with a large percentage of censored data points (i.e, data points where an exact time of age at onset is not known), censoring times were generated from a uniform distribution to achieve a random choice of censoring rate of 40% or 60%. To ensure valid inference, we used bootstrap resampling of families to compute the standard errors and construct confidence intervals for the estimators. The covariates included in the simulation models (sex of the relative and proband) were fixed at the same values as the real data. We did not include site of enrollment as a covariate in the simulation model since there was no significant heterogeneity between sites on the penetrance of PD in the real data analysis.

The goal of the simulation study is to examine the bias and efficiency of estimators. We evaluated the bias, empirical standard deviation, average of the estimated standard errors, and coverage probability corresponding to nominal 95% confidence intervals. We set the initial values to be zero in our simulations and data analysis and we did not find the algorithm to be sensitive to the choice of initial values. A reasonable way to select the initial values of parameters is to use baseline hazard ratios estimated from the probands data where all the genotypes are observed. We used exponential random variables with a mean of one to weight each observation. The simulation results from 1,000 replications with 1,000 bootstrap samples for each simulated data are given in Table 1 and Table 2.

The parameter estimates of the Cox proportional hazards model for the simulation study under both 40% and 60% censoring rates in Table 1 suggest that estimated hazard ratios (HR) are close to true HRs with small bias. The empirical variability agrees with the variance estimate based on the bootstrap and the coverage probabilities are close to the nominal level of 95%. Specifically, the estimated HR of predicted carrier male relatives to predicted non-carrier male relatives adjusted for other covariates was close to the true HR with bias < 0.01 (Estimated HR=1.52, True HR=1.52) and the estimated standard errors agrees adequately with the empirical standard deviation (SE=0.21, SD=0.21) with the coverage probability 94%. Moreover, higher HR was estimated for the predicted carrier female relatives compared to predicted non-carrier female relatives adjusted for other covariates which was close to the true HR (Estimated HR=5.20, True HR=4.99) and the estimated standard errors agrees adequately with the empirical standard deviation (SE=0.80, SD=0.83) with the coverage probability 93%. Our method performs well even under 60% censoring rate. As the censoring rate increases from 40% to 60%, we observed the empirical standard deviation and estimated standard errors increased slightly.

In Table 2, we present the average estimated values of the cumulative distribution functions ($\widehat{F}_1$ for carriers and $\widehat{F}_0$ for non-carriers) in male and female relatives at various ages with their performances. When cumulative risk in male and female relatives were examined separately, the small bias of estimated penetrance was observed through out the entire range of age (see Table 2 and Figure 1). Specifically, the bias of estimated penetrance to age 80 among predicted mutation carrier male relative with 40% censoring rate was -0.02% and the estimated standard errors agrees adequately with the empirical standard deviation (SE=4.19%, SD=4.28%), and the coverage probability of 93.4% was close to the nominal level. Similar results were observed in female relatives and our method performs well under both 40% and 60% censoring rates. We note that as the censoring rate increases, the bias and the variance estimates tend to increase and the coverage probability tends to decrease. This makes a wider 95% confidence interval for the cumulative risk estimates for 60% censoring rate compared to 40% censoring rate (see Figure 1 and Figure S2). In the Supplementary Material, we observed similar results for the overall penetrance estimates marginalized by relative's sex and proband's sex (see Table S1, Figure S1) as well as for the penetrance estimates in male and female relatives of male and female probands (see Table S2 and Figure S3 and S4).

Table 1: Summary results for the estimated hazard ratios in the simulation ($\times 10^{-2}$) with censoring rate of 40% or 60%.

| Censor | Variable | Parameter | True HR | Est HR | Bias | SD | SE | CP |
|---|---|---|---|---|---|---|---|---|
| 40% | Carrier status in male | $\beta + \theta$ | 1.52 | 1.52 | <0.01 | 0.21 | 0.21 | 0.94 |
| | Carrier status in female | $\beta$ | 4.99 | 5.20 | 0.20 | 0.83 | 0.80 | 0.93 |
| | Relative's sex in carriers | $\eta + \theta$ | 0.73 | 0.73 | <0.01 | 0.13 | 0.13 | 0.93 |
| | Relative's sex in non-carriers | $\eta$ | 2.39 | 2.44 | 0.05 | 0.18 | 0.18 | 0.93 |
| | Carriers · Relative's sex interaction | $\theta$ | 0.31 | 0.30 | <0.01 | 0.06 | 0.06 | 0.93 |
| | Proband's sex | $\gamma$ | 0.71 | 0.71 | <0.01 | 0.04 | 0.04 | 0.95 |
| 60% | Carrier status in male | $\beta + \theta$ | 1.52 | 1.49 | -0.03 | 0.24 | 0.24 | 0.96 |
| | Carrier status in female | $\beta$ | 4.99 | 4.84 | -0.15 | 0.89 | 0.94 | 0.95 |
| | Relative's sex in carriers | $\eta + \theta$ | 0.73 | 0.77 | 0.04 | 0.16 | 0.16 | 0.94 |
| | Relative's sex in non-carriers | $\eta$ | 2.39 | 2.43 | 0.04 | 0.22 | 0.23 | 0.95 |
| | Carriers · Relative's sex interaction | $\theta$ | 0.31 | 0.32 | 0.01 | 0.08 | 0.08 | 0.95 |
| | Proband's sex | $\gamma$ | 0.71 | 0.71 | 0.01 | 0.05 | 0.05 | 0.94 |

The Cox proportional hazards model can be written as

$\lambda(t|Carrier\,status\,X_i, Relative's\,sex_i, Proband's\,sex_i)$
$= \lambda_0(t)\exp\{\beta X_i + \eta I(Relative's\,sex_i = Male) + \theta X_i I(Relative's\,sex_i = Male) + \gamma I(Proband's\,sex_i = Male).$

Denote the true hazard ratio (True HR), estimated hazard ratio (Est HR), average estimation bias over 1,000 replications (Bias), empirical standard deviation (SD), average of estimated standard errors from bootstraps (SE), and coverage probability corresponding to nominal 95% confidence intervals (CP).

Table 2: Summary results for the penetrance estimates in male and female relatives marginalized by proband's sex in the simulation ($\times 10^{-2}$) with censoring rate of 40% or 60%.

| Censor | Relative's sex | Age | Carriers $\widehat{F}_1(\cdot)$ | | | | | Non-Carriers $\widehat{F}_0(\cdot)$ | | | | |
|---|---|---|---|---|---|---|---|---|---|---|---|---|
| | | | True Risk | Bias | SD | SE | CP | True Risk | Bias | SD | SE | CP |
| 40% | Male | 60 | 16.36 | <0.01 | 1.96 | 1.96 | 94.4 | 11.07 | 0.05 | 0.80 | 0.81 | 95.9 |
| | | 65 | 23.37 | 0.01 | 2.59 | 2.57 | 94.7 | 16.05 | 0.09 | 0.99 | 0.99 | 95.2 |
| | | 70 | 31.95 | -0.01 | 3.22 | 3.20 | 94.6 | 22.36 | 0.12 | 1.17 | 1.18 | 94.1 |
| | | 75 | 41.85 | -0.04 | 3.82 | 3.78 | 94.1 | 30.01 | 0.15 | 1.38 | 1.36 | 94.4 |
| | | 80 | 52.59 | -0.02 | 4.28 | 4.19 | 93.4 | 38.84 | 0.24 | 1.49 | 1.51 | 95.9 |
| | Female | 60 | 21.71 | 0.32 | 2.58 | 2.55 | 94.3 | 4.80 | -0.06 | 0.42 | 0.42 | 94.4 |
| | | 65 | 30.54 | 0.43 | 3.29 | 3.24 | 94.2 | 7.07 | -0.08 | 0.57 | 0.55 | 93.7 |
| | | 70 | 40.94 | 0.50 | 3.97 | 3.88 | 93.9 | 10.07 | -0.12 | 0.74 | 0.72 | 93.0 |
| | | 75 | 52.34 | 0.52 | 4.45 | 4.36 | 94.4 | 13.91 | -0.16 | 0.93 | 0.91 | 93.2 |
| | | 80 | 63.91 | 0.53 | 4.69 | 4.53 | 93.8 | 18.67 | -0.17 | 1.12 | 1.12 | 94.3 |
| 60% | Male | 60 | 16.36 | -0.14 | 2.20 | 2.18 | 93.5 | 11.07 | 0.17 | 0.84 | 0.85 | 95.0 |
| | | 65 | 23.37 | -0.21 | 2.91 | 2.88 | 94.0 | 16.05 | 0.23 | 1.04 | 1.05 | 95.2 |
| | | 70 | 31.95 | -0.26 | 3.63 | 3.62 | 94.2 | 22.36 | 0.34 | 1.29 | 1.27 | 94.2 |
| | | 75 | 41.85 | -0.51 | 4.26 | 4.31 | 94.7 | 30.01 | 0.30 | 1.48 | 1.49 | 94.6 |
| | | 80 | 52.59 | -0.98 | 4.80 | 4.83 | 94.3 | 38.84 | 0.06 | 1.76 | 1.73 | 94.1 |
| | Female | 60 | 21.71 | -0.72 | 2.79 | 2.85 | 94.1 | 4.80 | 0.03 | 0.48 | 0.48 | 94.4 |
| | | 65 | 30.54 | -0.99 | 3.62 | 3.68 | 94.2 | 7.07 | 0.04 | 0.65 | 0.65 | 94.9 |
| | | 70 | 40.94 | -1.21 | 4.40 | 4.50 | 94.5 | 10.07 | 0.07 | 0.85 | 0.85 | 94.6 |
| | | 75 | 52.34 | -1.61 | 5.02 | 5.13 | 94.3 | 13.91 | 0.03 | 1.07 | 1.08 | 94.6 |
| | | 80 | 63.91 | -2.14 | 5.38 | 5.46 | 93.2 | 18.67 | -0.13 | 1.33 | 1.33 | 94.1 |

Denote the true cumulative risk (True Risk), average estimation bias over 1,000 replications (Bias), empirical standard deviation (SD), average of estimated standard errors from bootstraps (SE), and coverage probability corresponding to nominal 95% confidence intervals (CP). Results are based on 1,000 simulations with sample size $n = 2266$.

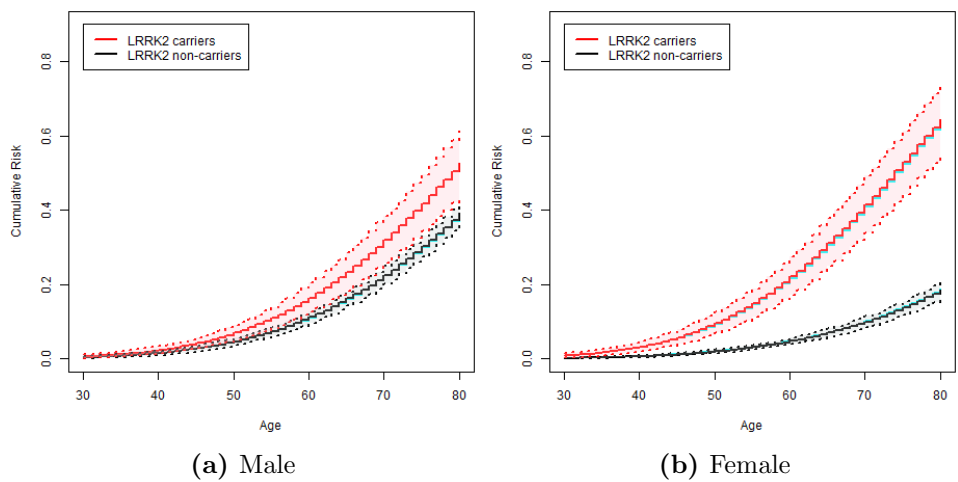

(a) Male  (b) Female

Figure 1: Estimated cumulative risk functions in the simulation with 40% censoring rate: Carriers (red solid line) and non-carriers (black solid line) in male and female relatives marginalized by proband's sex with their 95% confidence intervals (dashed lines) and true cumulative risk functions (blue solid line).

## 4. Application to the AJ Penetrance Study

Since mutations in the *LRRK2* gene are identified as a potential cause of autosomal dominant idiopathic Parkinson's disease (PD) [14], it is essential to estimate the cumulative risk of Parkinson's disease for LRRK2 mutation carriers for genetic counseling purposes [4, 9]. Ashkenazi Jews (AJ) are known to have high frequencies of G2019S mutations in *LRRK2* gene [19]. The risk for *LRRK2* G2019S mutation carriers vary widely in the literature [4]. Moreover, the penetrance estimates can be modified by genetic or environmental risk factors of age at onset or demographic factors including ethnic group or gender [5].

To provide precise risk prediction, we estimate the age-specific cumulative risk of Parkinson's disease for *LRRK2* G2019S carriers and non-carriers in the AJ cohort adjusted for multiple risk factors. Due to the low frequencies of *LRRK2* G2019S mutations in AJ population controls [20], we studied familial aggregaton of PD using the same validated family history interview at three academic centers specialized in the care of PD [5, 7, 19]. The family history data on the initial samples (probands) were collected from the Michael J. Fox Foundation Ashkenazi Jewish *LRRK2* Consortium [5, 9, 19]. Although the probands reported on family history of PD, most of the genotypes of the relatives were not observed [5]. The unobserved genotypes were inferred by the *LRRK2* G2019S mutation status in the probands using the kin-cohort method under a Cox hazards model to estimate the penetrance of *LRRK2* PD in first-degree relatives [5, 9]. The probands were excluded from the analysis to avoid ascertainment bias [9]. All probands were genotyped for the *LRRK2* G2019S mutation and glucocerebrosidase (GBA) mutations [5, 9]. To investigate the effect of *LRRK2* mutations on PD risk, we excluded probands carrying other known genetic risk factors such as GBA mutations [5, 9].

The data consists of 2266 first-degree relatives (i.e., 727 parents, 575 siblings, and 964 children) from 474 families. The participants were recruited at three sites: Beth Israel Medical Center (n=136), Columbia University Medical Center (n=146), and Tel-Aviv Medical Center (n=192). The prevalence of *LRRK2* G2019S mutation was estimated to be 2% in the AJ population [20]. There were four groups of mutation probabilities of relatives, $p_g \in \{0, 0.02, 0.51, 1\}$ [9], with frequencies 3%, 71%, 22%, and 4%, respectively. Hence, 93% of the first-degree relatives were not genotyped. Specifically, the percentage of missing genotypes was similar in siblings and children (91%) and slightly higher in parents (98%). There were 127 relatives with PD (5.6%) and relatives with censored age-at-onset were excluded from the analysis.

The potential risk factors that may have an impact on the penetrance of PD were considered to improve the accuracy of the estimation. For example, the recent study by Marder et al. (2015) showed that relative's sex may modify the penetrance of *LRRK2* G2019S in AJ cohort [5]. In another study of Ashkenazi PD, the first-degree relatives of female probands with PD were more likely to have PD compared to that of male probands with PD, even after accounting for *LRRK2* G2019S mutations [21]. Moreover, the issue of ascertainment through different academic centers (site) may have an impact on the penetrance estimates due to the heterogeneity of samples. Therefore, we aim to estimate the age-specific cumulative risk of PD in *LRRK2* mutation carriers and non-carriers adjusted for relative's sex and carrier status interaction, proband's sex, and site of enrollment in the Ashkenazi Jewish *LRRK2* Consortium study. We performed bootstrap based on families, which accommodates correlation among relatives in the same family. We used exponential random variables with a mean of one and weighted each family to account for the differential family data.

We considered using B-splines to estimate $\beta(t)$ with the number of knots ranging from zero to three and degrees ranging from one to three. The location of each interior knot was evenly distributed at the quantiles. No penalty was introduced as the number of knots is small. The number of knots can be chosen by the Bayesian information criterion (BIC). In the *LRRK2* study, among models with a time-varying $\beta(t)$, the simplest model with no knots had the smallest BIC (see

Table S4). Moreover, the model assuming a time-invariant genotype effect $\beta$ further had a smaller BIC. Therefore, we fit a final model with a time-invariant hazard for $LRRK2$ mutation, which maintains parsimony and facilitates easy interpretation. We also examined the Cox proportional hazards assumption using the probands data and the assumption was not rejected.

Marginal cumulative risk estimates of PD obtained using (5) based on 1,000 bootstrap resampling of families are shown in Table S3 along with 95% confidence intervals. After adjusting for relative's sex by $LRRK2$ G2019S mutation interaction, proband's sex, and recruitment sites, the penetrance of PD in relatives predicted to carry a G2019S mutation was 24.8% (95% CI: $16.3 - 34.3\%$) to age 80, whereas the cumulative risk for predicted non-carriers was 10.9% (95% CI: $8.1 - 14.5\%$) to age 80. The penetrance estimates were slightly higher for carriers and slightly lower for non-carriers both by 1% than when unadjusted for the covariates, reported in Marder et al. (2015), but these changes remained nearly identical.

Figure S5 shows the estimated marginal cumulative risk of PD where the predicted carrier relatives had a dramatic increase in the risk of developing PD after age 60 as compared to a lower increase in the predicted non-carrier relatives. While the study unadjusted for the covariates [5] reported the penetrance of PD in relatives predicted to carry a G2019S mutation was almost 3 fold higher than non-carrier relatives (HR=2.89; 95% CI: $1.73 - 4.55$; $p < 0.001$), when we examined the interaction effect between the mutation carrier status and relative's sex on the risk of PD, the effect of the carrier status on the risk of PD was significantly modified by gender (HR of interaction effect $\theta = 0.3; p = 0.03$) such that the effect was reduced for the male relatives (HR=1.47) and elevated for the female relatives (HR=4.95) (see Table 3). Specifically, when the male and female relatives were examined separately, the risk of PD for female relatives predicted to have a G2019S mutation was five-fold higher than non-carrier females (HR=4.95; 95% CI: $2.55 - 9.67$; $p < 0.001$) (Figure 2b), while for male relatives the risk of PD was not significantly increased among male relatives predicted to carry G2019S mutation compared to non-carrier males (HR=1.47; 95% CI: $0.53 - 3.07$; $p = 0.47$) (Figure 2a).

While the penetrance in G2019S carrier relatives and non-carrier relatives differed by gender, the penetrance to age 80 among predicted mutation carrier male relatives 21.4% (95% CI: $8.9-35.7\%$),

Table 3: Estimated hazard ratios of Parkinson's disease onset in the Ashkenazi Jewish $LRRK2$ Consortium study.

| Variable | Parameter | Est HR | Lower limit | Upper limit | $p$ value |
|---|---|---|---|---|---|
| Carrier status in male | $\beta + \theta$ | 1.47 | 0.53 | 3.07 | 0.472 |
| Carrier status in female | $\beta$ | 4.95 | 2.55 | 9.67 | <0.001 |
| Relative's sex in carriers | $\eta + \theta$ | 0.72 | 0.26 | 1.56 | 0.410 |
| Relative's sex in non-carriers | $\eta$ | 2.43 | 1.50 | 4.06 | <0.001 |
| Carriers · Relative's sex interaction | $\theta$ | 0.30 | 0.09 | 0.86 | 0.028 |
| Probands's sex | $\gamma_1$ | 0.71 | 0.49 | 1.02 | 0.053 |
| Beth Israel vs Columbia | $\gamma_2$ | 0.78 | 0.46 | 1.28 | 0.328 |
| Tel Aviv vs Columbia | $\gamma_3$ | 1.18 | 0.72 | 1.97 | 0.483 |
| Beth Israel vs Tel Aviv | $\gamma_2 - \gamma_3$ | 0.66 | 0.38 | 1.11 | 0.113 |

The Cox proportional hazards model can be written as

$\lambda(t|Carrier\,status\,X_i, Relative's\,sex_i, Proband's\,sex_i, Beth\,Israel_i, Tel\,Aviv_i)$
$= \lambda_0(t)\exp\{\beta X_i + \eta^T I(Relative's\,sex_i = Male) + \theta^T X_i I(Relative's\,sex_i = Male) + \gamma_1^T I(Proband's\,sex_i = Male)$
$+ \gamma_2^T I(Site_i = Beth\,Israel) + \gamma_3^T I(Site_i = Tel\,Aviv)\}.$

where Beth Israel and Tel Aviv are dummy variables for three sites with Columbia as reference site.
Denote the estimated hazard ratio (Est HR), 95% confidence interval for the hazard ratio (Lower limit and Upper limit), and $p$ value.

was not statistically different from predicted carrier female relatives 28.2% (95% CI: $18.7 - 40.1\%$); HR male to female: 0.72 (95% CI: $0.26 - 1.56$; $p = 0.41$) (see Table 3 and 4). In contrast, the risk of PD to age 80 among predicted non-carrier male relatives 15.1% (95% CI: $10.8 - 20.5\%$), was higher than for predicted non-carrier female relatives 6.5% (95% CI: $4.0 - 10.2\%$); HR male to female 2.43 (95% CI: $1.50 - 4.06$; $p < 0.001$) was increased by 3% after the adjustment (HR male to female before the adjustment: 2.40; 95% CI: $1.50 - 4.15$; $p < 0.001$) [5]. The penetrance estimates were slightly lower for carrier male relatives by 1% and slightly higher for non-carrier male relatives by 0.1% than when unadjusted for the covariates, reported in Marder et al. (2015), but these changes remained almost identical. Therefore, the large gender difference of penetrance between carrier relatives and non-carrier relatives is due to the different PD distribution in non-carriers among

Table 4: Estimated cumulative risk of Parkinson's disease onset in $LRRK2$ carriers and non-carriers in male and female relatives marginalized by proband's sex and site of enrollment in the Ashkenazi Jewish $LRRK2$ Consortium study ($\times 10^{-2}$).

| Relative's sex | Age | Carriers $\widehat{F}_1(\cdot)$ | | | Non-Carriers $\widehat{F}_0(\cdot)$ | | |
|---|---|---|---|---|---|---|---|
| | | Risk | Lower limit | Upper limit | Risk | Lower limit | Upper limit |
| Male | 60 | 6.16 | 2.31 | 11.22 | 4.23 | 2.55 | 6.44 |
| | 65 | 9.73 | 3.86 | 16.80 | 6.72 | 4.48 | 9.70 |
| | 70 | 14.32 | 5.68 | 24.50 | 9.98 | 6.85 | 14.16 |
| | 75 | 17.88 | 7.07 | 30.78 | 12.55 | 8.93 | 17.13 |
| | 80 | 21.37 | 8.88 | 35.67 | 15.09 | 10.80 | 20.50 |
| Female | 60 | 8.41 | 4.41 | 13.89 | 1.77 | 1.02 | 2.85 |
| | 65 | 13.20 | 7.64 | 20.36 | 2.83 | 1.64 | 4.54 |
| | 70 | 19.23 | 12.38 | 27.82 | 4.25 | 2.51 | 6.65 |
| | 75 | 23.82 | 15.35 | 34.22 | 5.39 | 3.33 | 8.25 |
| | 80 | 28.23 | 18.66 | 40.06 | 6.54 | 4.00 | 10.18 |

Denote the estimated age-specific risk (Risk) and 95% confidence intervals (Lower limit and Upper limit).

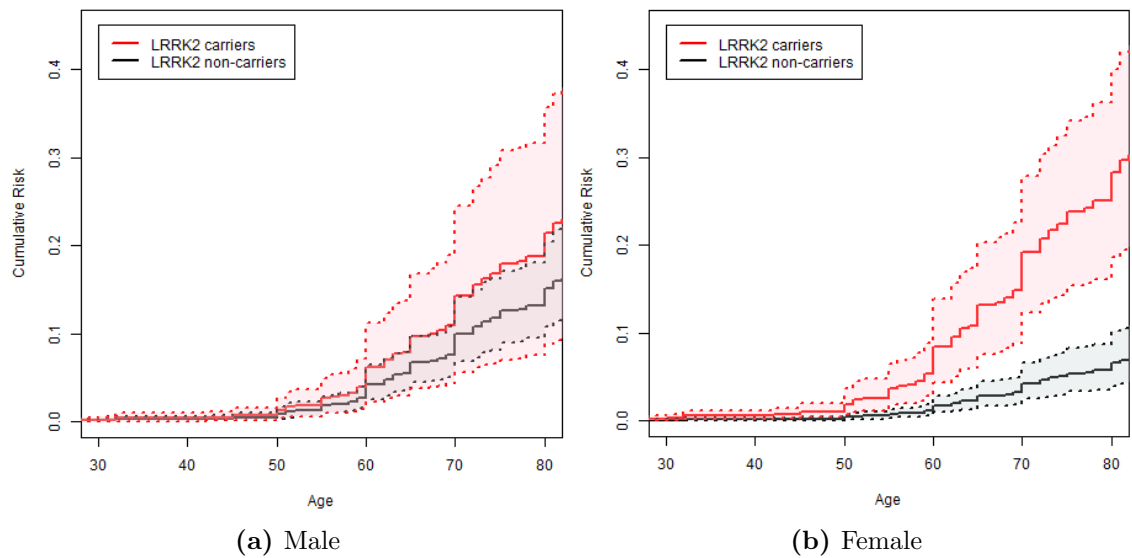

(a) Male  (b) Female

Figure 2: Estimated age-specific risk of Parkinson's disease in $LRRK2$ G2019S carriers (red solid line) and non-carriers (black solid line) in male and female relatives marginalized by proband's sex and site of enrollment with their 95% confidence intervals (dashed lines).

male and female relatives [5]. These HR are similar to the previous findings where the proband's sex and site of enrollment were not controlled additionally [5].

When we specifically examined the effect of proband's sex and site of enrollment on the penetrance of PD, the risk of PD was not significantly reduced among male probands compared to female probands (HR=0.71; 95% CI: $0.49-1.02$; $p=0.05$) and there was no significant heterogeneity between sites on the risk of PD, adjusted for other covariates; HR Beth Israel to Columbia=0.78 (95% CI: $0.46-1.28$; $p=0.33$), HR Tel Aviv to Columbia=1.18 (95% CI: $0.72-1.97$; $p=0.48$), HR Beth Israel to Tel Aviv =0.66 (95% CI: $0.38-1.11$; $p=0.11$) (see Table 3). This implies that proband's sex and site of enrollment do not play a significant role on the penetrance of PD for the data collected in the AJ *LRRK2* consortium.

To obtain the marginal risk distribution as in Section 2.3 to serve as design parameters, note that since the mutation status $X$ of the first-degree relatives is independent of covariates $\boldsymbol{W}_i$ collected on the relative (e.g., relative gender) and proband's covariates $\boldsymbol{Z}_i$ (e.g., proband's sex and enrollment site), $F(\boldsymbol{W}, \boldsymbol{Z}|X) = F(\boldsymbol{W}, \boldsymbol{Z})$ can be obtained nonparametrically based on the observed data. When the independence assumption does not hold for other applications, the conditional distribution of $(\boldsymbol{W}, \boldsymbol{Z})$ given $X$ can be estimated by a similar EM algorithm treating $X$ as missing data. The estimated penetrance for each relative's sex by carrier status marginalized by proband's sex and site of enrollment is reported in Table 4.

To assist with clinical trial planning, we show how to use the estimated penetrance function to design a disease modifying clinical trial of PD. In a study of Huntington's disease, sample size calculations used the length of abnormal expansion of CAG repeat in the *HTT* gene since the penetrance is near complete [22]. However, in the PD literatures, using estimated mutation risk to design a future clinical trial is less common due to the low penetrance for most mutations. The increased PD risk in *LRRK2* carriers compared to non-carriers in the AJ population provides a unique opportunity to power a future clinical trial in this population.

Assume a study recruits *LRRK2* carrier and non-carrier relatives from probands in the three sites in the AJ consortium. Assume that the PD risk in the placebo arm is the same as observed in our carrier relatives' data. We can obtain the placebo arm as the probability of developing PD within 5 years for *LRRK2* carrier relatives who are not diagnosed at the baseline as $p_0 = P(T \leq t+5|T>t, X=1)$ for any given baseline age $t$ using the marginalized estimators provided by our analysis. Assume the effect of the intervention is to reduce the PD risk in carriers to that as observed in the non-carrier relatives. Under this assumption, we can estimate the intervention arm as the proportion of subjects affected by the PD by the end of 5 years as $p_1 = P(T \leq t+5|T>t, X=0)$ for any given baseline age $t$. Then the sample size to achieve a power of 80% for testing the difference between placebo arm and intervention arm, $H_0: p_0 = p_1$, can be estimated.

Table 5 summarizes sample size information by baseline age. The results show that when recruiting asymptomatic *LRRK2* carriers at age 65, the risk of developing PD within next 5 years is 6.0% in the control arm. When assuming the intervention reduces the PD risk to the same as observed in the non-carrier group (scenario 1), the risk of developing PD in the intervention arm, is 2.5%. To detect a difference of 3.5% (1.1%-5.9%) at 80% power, the required sample size is $n=521$ per arm. The sample size needed for recruiting asymptomatic subjects at other age is also presented. We also estimated the sample size to achieve a power of 80% for testing a smaller risk difference between placebo and intervention arm (50% of the previous risk difference); that is, to test $H_0: p_0 = \widetilde{p}_1$, by letting $\widetilde{p}_1 = p_0 + (p_1-p_0)/2$. The results in Table 5 show that when recruiting pre-symptomatic *LRRK2* carriers at age 65, the required sample size to detect a difference of 1.7% at 80% power is $n=2,492$ per arm. A large sample size is required for a prevention trial under the specified design parameters due to small risk differences.

Table 5: Sample size estimates for hypothesis testing to achieve 80% power ($\alpha = 0.05$): Test difference between placebo arm and intervention arm (Scenario 1), test half of the risk reduction in scenario 1 (Scenario 2), probability of developing PD within 5 years in placebo arm ($p_0$, assumed to be the same as observed in carriers), intervention arm of scenario 1 ($p_1$, assumed to be the same as in non-carriers), intervention arm of scenario 2 ($\widetilde{p}_1$ that makes half of the risk reduction in scenario 1), difference between two arms (diff), and sample size $n$ per arm.

|     |       | Scenario 1 | | | Scenario 2 | | |
| --- | ----- | ----- | ----- | ----- | ----- | ----- | ----- |
| Age | $p_0$ | $p_1$ | diff  | $n$ per arm | $\widetilde{p}_1$ | diff | $n$ per arm |
| 60  | 0.045 | 0.018 | 0.027 | 657 | 0.032 | 0.013 | 3169 |
| 65  | 0.060 | 0.025 | 0.035 | 521 | 0.042 | 0.017 | 2492 |
| 70  | 0.049 | 0.020 | 0.029 | 622 | 0.034 | 0.014 | 2987 |
| 75  | 0.050 | 0.020 | 0.030 | 589 | 0.035 | 0.015 | 2839 |
| 80  | 0.088 | 0.037 | 0.051 | 354 | 0.062 | 0.026 | 1680 |

The estimated risk in placebo arm and two intervention arms are obtained by
$$p_0 = \hat{P}(T \leq t+5 | T > t, X = 1) = \left[\hat{F}_1(t+5) - \hat{F}_1(t)\right] / \left[1 - \hat{F}_1(t)\right],$$
$$p_1 = \hat{P}(T \leq t+5 | T > t, X = 0) = \left[\hat{F}_0(t+5) - \hat{F}_0(t)\right] / \left[1 - \hat{F}_0(t)\right],$$
$$\widetilde{p}_1 = p_0 + (p_1 - p_0)/2,$$
where $\hat{F}_1$ and $\hat{F}_0$ are the estimated marginal cumulative risk functions of PD-onset in *LRRK2* carriers and non-carriers, respectively (see Table S3).

## 5. Discussion

We propose a sieve maximum likelihood estimation method that permits adjustment for multiple covariates and interaction effects to estimate disease risk associated with genetic mutation in censored mixture data. The method allows more precise risk prediction by controlling for individual characteristics such as sex, ethnicity or other demographics. Moreover, gene-gene interactions and gene-environment interactions can also be handled within the framework of a single risk model using our method. These analyses may provide insights into whether these factors play a role in modifying penetrance. Previous method [9] has been proposed without considering covariates. Our model and method here permit more precise risk prediction by controlling for covariates that can further incorporate interaction effects with missing genotypes in relatives. In the application, when we examined the interaction effect between the mutation carrier status and relative's sex on the risk of PD, the effect of the carrier status on the risk of PD was significantly modified by gender ($p = 0.023$; Table 3). The proposed method can also be readily generalized to include time-dependent covariates. For example, PD risk can be associated with clinical or environmental time-varying covariates. For example, studies have reported that cigaratte smoking is inversely associated with PD [23]. Therefore, taking into account of duration of smoking or time since quitting is important in estimating penetrance [24]. Additionally, when the dimension of $\boldsymbol{Z}_i$ is high, it may be of interest to perform variable selection in the M-step of the EM algorithm. Furthermore, we included effect size calculation to power a clinical trial. The penetrance can be used to design a prevention trial in order to test an intervention to reduce the risk of PD in asymptomatic *LRRK2* p.G2019S mutation carrier relatives in a AJ population. The penetrance estimates can be used in genetic counseling setting. For example, female Ashkenazi Jewish relatives carrying *LRRK2* mutation with female probands recruited at Beth Israel are estimated to have 27.19% cumulative risk of PD by age 80 in our application.

As an extension of our method, under certain assumptions we can calibrate the genetic risk prediction model with covariates from samples to the population. Currently, we estimated the baseline hazard function from the study sample. However, the sample baseline hazard rate may not reflect the population baseline hazard rate. Instead, we can estimate the baseline rate in non-carriers from an external source and include in the model to obtain cumulative risk distribution in carriers under certain assumptions. When the population prevalence of deleterious mutation is low, the baseline cumulative hazard of a disease is readily available from national-based surveys or administrative databases. Assuming the hazard ratios for $(X_i, \boldsymbol{W}_i, \boldsymbol{Z}_i)$ are the same between the study sample and the underlying population, the model for the study sample is $\lambda_s(t|X_i, \boldsymbol{W}_i, \boldsymbol{Z}_i) = \lambda_{0s}(t)\exp\{\beta(t)X_i + \boldsymbol{\eta}^T\boldsymbol{W}_i + \boldsymbol{\theta}^T\boldsymbol{W}_i\boldsymbol{Z}_i + \boldsymbol{\gamma}^T\boldsymbol{Z}_i\}$, and for the population is $\widetilde{\lambda}(t|X_i, \boldsymbol{W}_i, \boldsymbol{Z}_i) = \widetilde{\lambda}_0(t)\exp\{\beta(t)X_i + \boldsymbol{\eta}^T\boldsymbol{W}_i + \boldsymbol{\theta}^T\boldsymbol{W}_iX_i + \boldsymbol{\gamma}^T\boldsymbol{Z}_i\}$, where the difference between two models is the baseline hazard function. Combining the baseline cumulative hazard function obtained from external resource with regression coefficients estimated from the study sample, the genetic risk function for the carrier group can be obtained under specified assumptions. Such estimation can be performed when planning clinical trials on certain populations with known baseline cumulative risk of disease.

In the studies we considered here, relatives' disease onset phenotypes were collected cross-sectionally from examining probands in each family. Probands' ascertainment is through a disease consortium and primarily clinic-based. Thus, relatives' missing genotypes due to death and censoring is unlikely to be directly associated with their disease onset ages given that a proband is recruited. In other words, relatives' missing genotypes or censoring can be conditionally independent of their PD onset given proband's ascertainment scheme and covariates, and hence properly accounted for. One source of bias may be length-biased sampling (for example, probands who are younger and with longer duration of PD may be more likely to be recruited), and alternative methods [25] can be considered in this case. Our method can be modified in the presence of competing risks to account for cause-specific hazards along the lines of [26]. Unfortunately, a limitation of *LRRK2* study is that the cause of death in deceased relatives was not available. Lastly, due to a small number of GBA carriers in the *LRRK2* study sample, estimation of *LRRK2* by *GBA* interaction was not stable and thus not reported.

## Acknowledgements

This work was supported by the Michael J. Fox Foundation, and A. Lee receives research support from the NIH TL1 Personalized Medicine Training Program (TL1TR000082). K. Marder receives research support from the NIH: NS036630, 1UL1 RR024156-01, PO412196-G, and PO412196-G. R. Alcalay receives research support from the NIH (K02NS080915), the Parkinson's Disease Foundation, the Smart Foundation, and the Michael J. Fox Foundation. A. Orr-Urtreger receives research support from the Kahn Foundation, Chief Scientist of the Israeli Ministry of Health, and the Michael J. Fox Foundation for Parkinson's Research. N. Giladi serves as a member of the Editorial Board for the Journal of Parkinson's Disease. N. Giladi received research support from the Michael J Fox Foundation, the National Parkinson Foundation, the European Union 7th Framework Program and the Israel Science Foundation as well as from Teva NNE program, LTI, and Abviee and CHDI. S. Bressman has received research support from the Michael J. Fox Foundation, NIH, and Dystonia Medical Research Foundation. Y. Wang receives research support from the NIH (NS073671, NS082062) and the Michael J. Fox Foundation.

Supporting Information

Additional supporting information including additional simulations and data analyses results may be found in the online version of this article at the publishers web site.

Appendix A. Additional simulation results

Based on the simulation results from 1,000 replications with 1,000 bootstrap samples in Section 3 and using model (5), we estimated the marginal cumulative risk functions in carriers and non-carriers in Figure S1 and the performance of the estimates at various ages with censoring rate of 40% or 60% in Table S1. The penetrance estimates in male and female relatives were examined separately along with their 95% confidence intervals with the censoring rate of 40% and 60% in Figure 1 and Figure S2, respectively. Moreover, Figure S3 and S4 presents the penetrance estimates in male and female relatives of male and female probands and Table S2 shows the performance of the estimates at various ages with the censoring rate of 40% or 60%.

In Table 2, S1, and S2, we observed the small bias of estimated penetrance over the entire range of age and the estimated standard errors agrees adequately with the empirical standard deviations, and the coverage probabilities were close to the nominal level.

Table S1: Summary results for the marginal penetrance estimates marginalized by relative's sex and proband's sex in the simulation ($\times 10^{-2}$) with censoring rate of 40% or 60%.

|        |     | Carrier $\widehat{F}_1(\cdot)$ |       |      |      |      | Non-Carrier $\widehat{F}_0(\cdot)$ |        |      |      |      |
|--------|-----|-----------|-------|------|------|------|-----------|--------|------|------|------|
| Censor | Age | True Risk | Bias  | SD   | SE   | CP   | True Risk | Bias   | SD   | SE   | CP   |
| 40%    | 60  | 19.00     | 0.16  | 1.69 | 1.71 | 94.4 | 7.97      | <0.01  | 0.56 | 0.56 | 95.3 |
|        | 65  | 26.91     | 0.21  | 2.13 | 2.13 | 94.0 | 11.61     | <0.01  | 0.70 | 0.69 | 94.6 |
|        | 70  | 36.38     | 0.24  | 2.56 | 2.56 | 94.4 | 16.29     | <0.01  | 0.83 | 0.82 | 95.1 |
|        | 75  | 47.02     | 0.24  | 2.91 | 2.90 | 94.5 | 22.06     | <0.01  | 0.97 | 0.95 | 94.3 |
|        | 80  | 58.16     | 0.25  | 3.19 | 3.11 | 94.1 | 28.88     | 0.04   | 1.04 | 1.06 | 96.1 |
| 60%    | 60  | 19.00     | -0.43 | 1.81 | 1.87 | 94.8 | 7.97      | 0.10   | 0.59 | 0.58 | 94.4 |
|        | 65  | 26.91     | -0.60 | 2.32 | 2.38 | 94.4 | 11.61     | 0.14   | 0.72 | 0.72 | 95.6 |
|        | 70  | 36.38     | -0.73 | 2.80 | 2.90 | 95.3 | 16.29     | 0.21   | 0.89 | 0.86 | 94.2 |
|        | 75  | 47.02     | -1.05 | 3.20 | 3.37 | 95.7 | 22.06     | 0.17   | 1.01 | 1.01 | 94.5 |
|        | 80  | 58.16     | -1.55 | 3.56 | 3.70 | 94.2 | 28.88     | -0.03  | 1.19 | 1.16 | 93.9 |

Denote the true cumulative risk (True Risk), average estimation bias over 1,000 replications (Bias), empirical standard deviation (SD), average of estimated standard errors from bootstraps (SE), and coverage probability corresponding to nominal 95% confidence intervals (CP). Results are based on 1,000 simulations with sample size $n = 2266$.

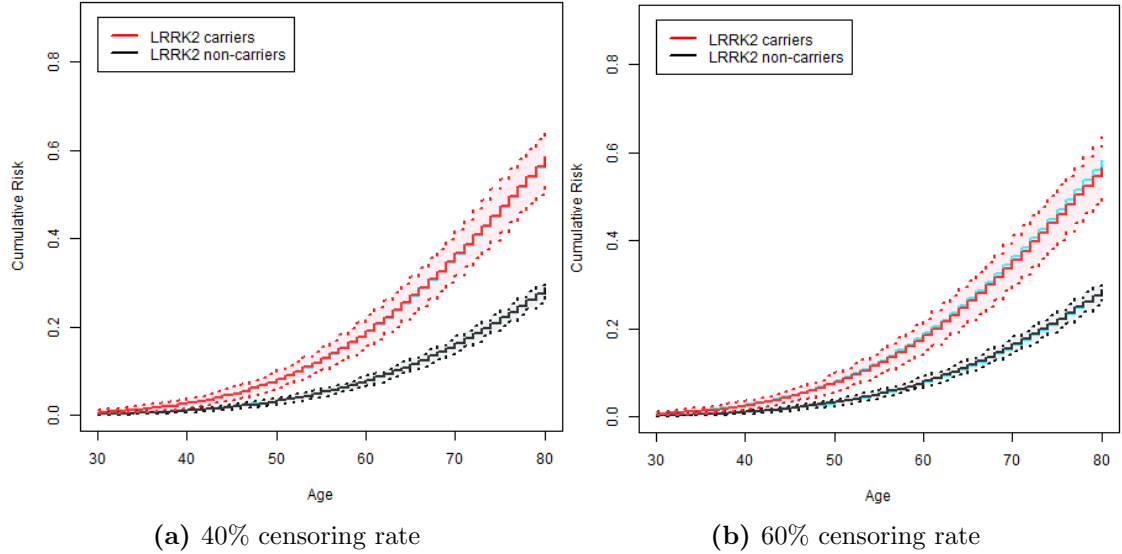

(a) 40% censoring rate  (b) 60% censoring rate

Figure S1: Estimated marginal cumulative risk functions in the simulation with censoring rate of 40% or 60%: Carriers (red solid line) and non-carriers (black solid line) with their 95% confidence intervals (dashed lines) and true cumulative risk functions (blue solid line).

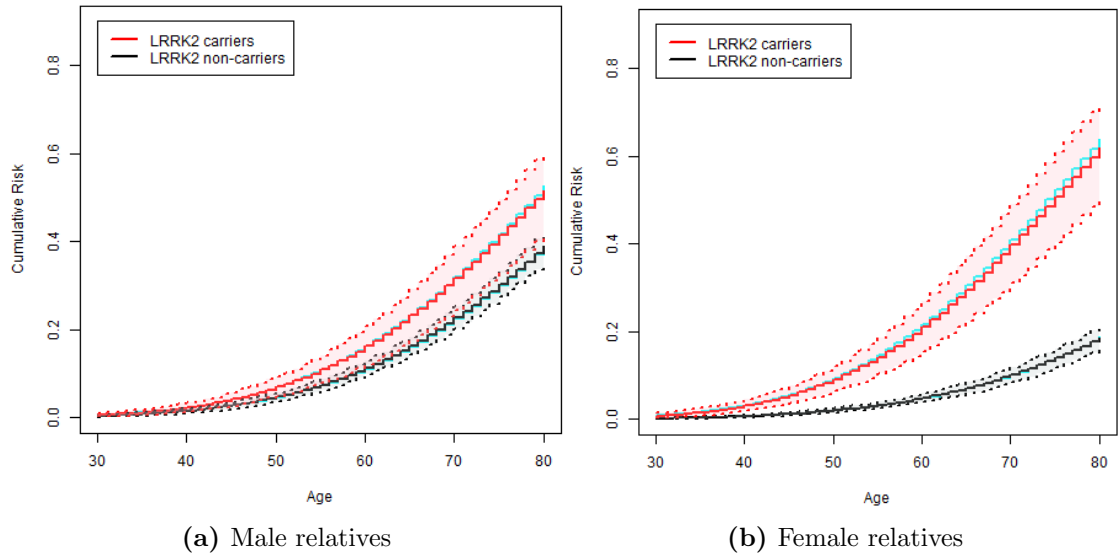

(a) Male relatives  (b) Female relatives

Figure S2: Estimated cumulative risk functions in the simulation with 60% censoring rate: Carriers (red solid line) and non-carriers (black solid line) in male and female relatives marginalized by probands' sex with their 95% confidence intervals (dashed lines) and true cumulative risk functions (blue solid line).

Table S2: Summary results for the penetrance estimates in male and female relatives of male and female probands in the simulation ($\times 10^{-2}$) with censoring rate of 40% or 60%.

| Censor | Rel sex | Prob sex | Age | Carrier $\widehat{F}_1(\cdot)$ ||||| Non-Carrier $\widehat{F}_0(\cdot)$ |||||
|---|---|---|---|---|---|---|---|---|---|---|---|---|---|
| | | | | True Risk | Bias | SD | SE | CP | True Risk | Bias | SD | SE | CP |
| 40% | Male | Male | 60 | 14.20 | 0.02 | 1.77 | 1.77 | 94.8 | 9.56 | 0.06 | 0.76 | 0.75 | 94.8 |
| | | | 65 | 20.43 | 0.04 | 2.38 | 2.35 | 94.5 | 13.92 | 0.10 | 0.96 | 0.95 | 94.1 |
| | | | 70 | 28.18 | 0.04 | 3.03 | 2.99 | 93.7 | 19.52 | 0.14 | 1.19 | 1.16 | 94.9 |
| | | | 75 | 37.33 | 0.03 | 3.68 | 3.62 | 94.0 | 26.41 | 0.18 | 1.44 | 1.39 | 93.6 |
| | | | 80 | 47.55 | 0.06 | 4.25 | 4.14 | 93.5 | 34.52 | 0.27 | 1.62 | 1.60 | 94.9 |
| | | Female | 60 | 19.49 | -0.04 | 2.36 | 2.36 | 94.3 | 13.26 | 0.04 | 1.01 | 1.03 | 94.6 |
| | | | 65 | 27.63 | -0.04 | 3.06 | 3.05 | 94.4 | 19.12 | 0.07 | 1.26 | 1.27 | 95.3 |
| | | | 70 | 37.40 | -0.08 | 3.73 | 3.73 | 94.7 | 26.47 | 0.10 | 1.49 | 1.53 | 95.7 |
| | | | 75 | 48.39 | -0.13 | 4.30 | 4.27 | 94.3 | 35.21 | 0.12 | 1.75 | 1.78 | 95.9 |
| | | | 80 | 59.88 | -0.13 | 4.63 | 4.55 | 93.7 | 45.08 | 0.20 | 1.90 | 1.97 | 96.1 |
| | Female | Male | 60 | 18.94 | 0.32 | 2.34 | 2.32 | 94.2 | 4.12 | -0.05 | 0.39 | 0.39 | 93.9 |
| | | | 65 | 26.90 | 0.44 | 3.06 | 3.01 | 94.1 | 6.08 | -0.06 | 0.53 | 0.52 | 94.2 |
| | | | 70 | 36.48 | 0.53 | 3.80 | 3.71 | 93.9 | 8.69 | -0.09 | 0.70 | 0.68 | 92.7 |
| | | | 75 | 47.31 | 0.58 | 4.40 | 4.30 | 94.5 | 12.04 | -0.12 | 0.90 | 0.87 | 93.5 |
| | | | 80 | 58.72 | 0.63 | 4.82 | 4.65 | 93.6 | 16.24 | -0.13 | 1.10 | 1.09 | 94.1 |
| | | Female | 60 | 25.71 | 0.33 | 3.07 | 3.03 | 93.9 | 5.78 | -0.08 | 0.53 | 0.52 | 94.0 |
| | | | 65 | 35.82 | 0.42 | 3.84 | 3.78 | 93.9 | 8.50 | -0.11 | 0.71 | 0.69 | 93.1 |
| | | | 70 | 47.39 | 0.46 | 4.48 | 4.40 | 94.2 | 12.07 | -0.16 | 0.92 | 0.90 | 93.5 |
| | | | 75 | 59.62 | 0.42 | 4.81 | 4.73 | 94.3 | 16.61 | -0.21 | 1.17 | 1.14 | 93.5 |
| | | | 80 | 71.42 | 0.38 | 4.78 | 4.63 | 94.1 | 22.18 | -0.24 | 1.41 | 1.40 | 93.8 |
| 60% | Male | Male | 60 | 14.20 | -0.08 | 2.02 | 1.97 | 93.7 | 9.56 | 0.17 | 0.81 | 0.81 | 94.7 |
| | | | 65 | 20.43 | -0.14 | 2.71 | 2.64 | 93.5 | 13.92 | 0.24 | 1.04 | 1.03 | 94.9 |
| | | | 70 | 28.18 | -0.17 | 3.46 | 3.39 | 94.3 | 19.52 | 0.35 | 1.32 | 1.29 | 93.7 |
| | | | 75 | 37.33 | -0.39 | 4.16 | 4.13 | 95.0 | 26.41 | 0.33 | 1.57 | 1.56 | 94.2 |
| | | | 80 | 47.55 | -0.83 | 4.81 | 4.77 | 94.3 | 34.52 | 0.13 | 1.88 | 1.84 | 94.7 |
| | | Female | 60 | 19.49 | -0.22 | 2.61 | 2.64 | 94.3 | 13.26 | 0.17 | 1.08 | 1.11 | 95.1 |
| | | | 65 | 27.63 | -0.33 | 3.41 | 3.43 | 94.5 | 19.12 | 0.22 | 1.37 | 1.39 | 95.2 |
| | | | 70 | 37.40 | -0.40 | 4.18 | 4.23 | 94.8 | 26.47 | 0.33 | 1.71 | 1.70 | 95.1 |
| | | | 75 | 48.39 | -0.69 | 4.77 | 4.90 | 95.2 | 35.21 | 0.26 | 1.98 | 2.02 | 95.1 |
| | | | 80 | 59.88 | -1.19 | 5.17 | 5.30 | 95.1 | 45.08 | -0.03 | 2.33 | 2.32 | 94.6 |
| | Female | Male | 60 | 18.94 | -0.60 | 2.56 | 2.60 | 93.8 | 4.12 | 0.04 | 0.45 | 0.45 | 95.0 |
| | | | 65 | 26.90 | -0.83 | 3.38 | 3.42 | 94.3 | 6.08 | 0.05 | 0.61 | 0.61 | 95.1 |
| | | | 70 | 36.48 | -1.04 | 4.22 | 4.28 | 94.3 | 8.69 | 0.08 | 0.82 | 0.81 | 95.5 |
| | | | 75 | 47.31 | -1.44 | 4.96 | 5.03 | 94.0 | 12.04 | 0.05 | 1.04 | 1.05 | 95.7 |
| | | | 80 | 58.72 | -2.01 | 5.50 | 5.55 | 94.0 | 16.24 | -0.08 | 1.30 | 1.31 | 96.1 |
| | | Female | 60 | 25.71 | -0.90 | 3.32 | 3.42 | 94.3 | 5.78 | 0.02 | 0.60 | 0.60 | 94.2 |
| | | | 65 | 35.82 | -1.22 | 4.22 | 4.32 | 93.9 | 8.50 | 0.03 | 0.82 | 0.81 | 93.8 |
| | | | 70 | 47.39 | -1.46 | 5.00 | 5.12 | 94.0 | 12.07 | 0.05 | 1.08 | 1.07 | 94.1 |
| | | | 75 | 59.62 | -1.86 | 5.49 | 5.62 | 93.8 | 16.61 | -0.01 | 1.37 | 1.37 | 94.5 |
| | | | 80 | 71.42 | -2.32 | 5.58 | 5.68 | 93.8 | 22.18 | -0.21 | 1.70 | 1.69 | 94.1 |

Denote the true cumulative risk (True Risk), average estimation bias over 1,000 replications (Bias), empirical standard deviation (SD), average of estimated standard errors from bootstraps (SE), and coverage probability corresponding to nominal 95% confidence intervals (CP). Results are based on 1,000 simulations with sample size $n = 2266$.

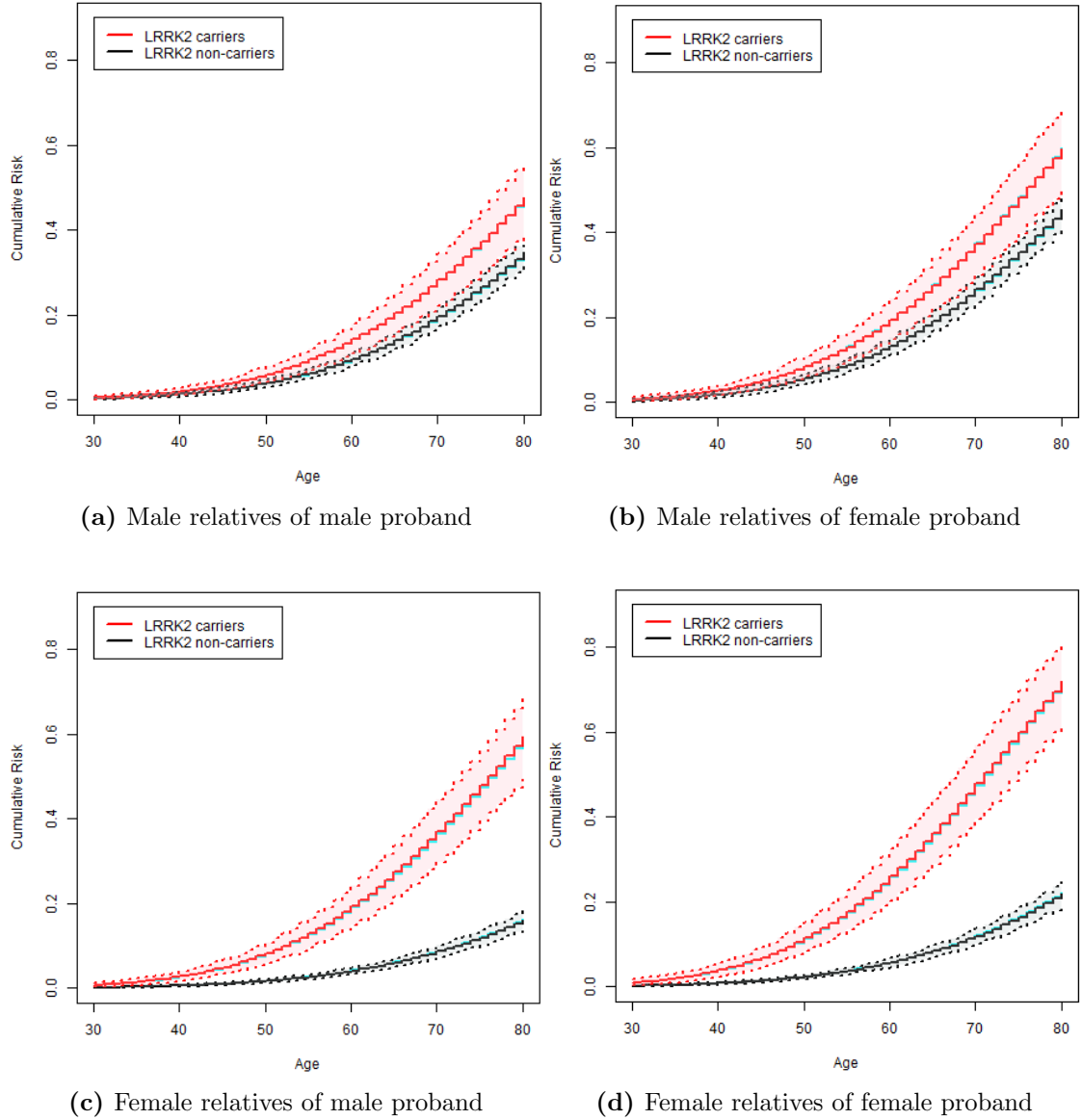

Figure S3: Estimated cumulative risk functions in the simulation with 40% censoring rate: Carriers (red solid line) and non-carriers (black solid line) in male and female relatives of male and female probands with their 95% confidence intervals (dashed lines) and true cumulative risk functions (blue solid line).

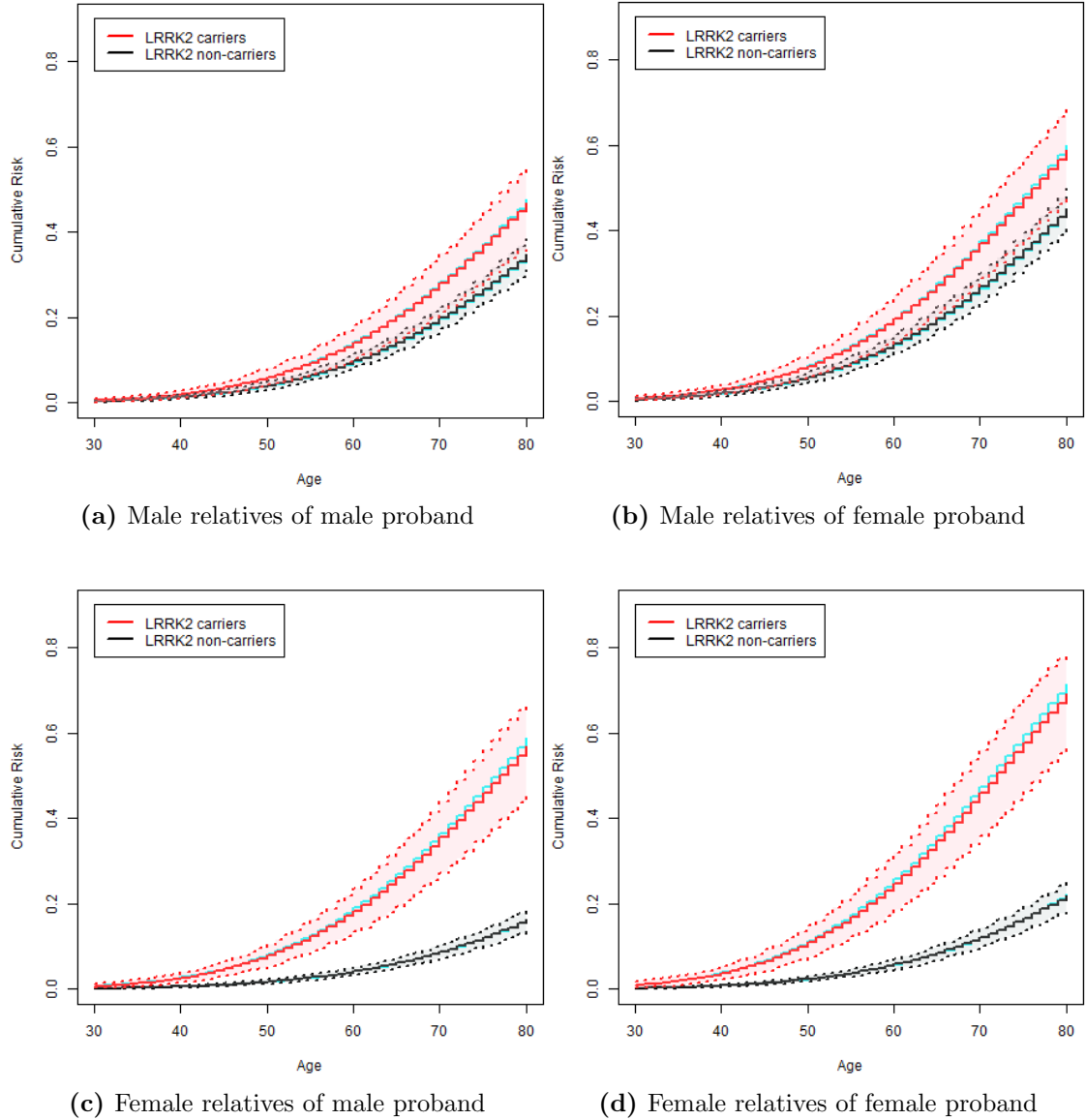

Figure S4: Estimated cumulative risk functions in the simulation with 60% censoring rate: Carriers (red solid line) and non-carriers (black solid line) in male and female relatives of male and female probands with their 95% confidence intervals (dashed lines) and true cumulative risk functions (blue solid line).

Appendix B. Additional data analyses results

Based on the parameter estimates that we obtained in the Ashkenazi Jewish *LRRK2* Consortium study in Section 4, the marginal cumulative risk functions in *LRRK2* G2019S carriers and non-carriers can be estimated using model (5) and the results are in Table S3 and Figure S5. We adjusted for relative's sex and carrier status interaction, proband's sex, and site of enrollment to provide precise risk prediction and we compared the penetrance estimates to the one in Marder et al. (2015) where none of the covariates were controlled.

Table S3: Estimated marginal cumulative risk of Parkinson's disease onset in *LRRK2* carriers and non-carriers in the Ashkenazi Jewish *LRRK2* Consortium study ($\times 10^{-2}$).

|  | Age | Carrier $\widehat{F}_1(\cdot)$ | | | Non-Carrier $\widehat{F}_0(\cdot)$ | | |
| --- | --- | --- | --- | --- | --- | --- | --- |
|  |  | Risk | Lower limit | Upper limit | Risk | Lower limit | Upper limit |
| Adjusted (Proposed) | 60 | 7.27 | 4.18 | 11.16 | 3.01 | 1.94 | 4.45 |
|  | 65 | 11.44 | 6.86 | 16.50 | 4.80 | 3.28 | 6.83 |
|  | 70 | 16.74 | 10.84 | 23.46 | 7.15 | 4.93 | 9.88 |
|  | 75 | 20.81 | 13.27 | 28.86 | 9.01 | 6.68 | 11.99 |
|  | 80 | 24.75 | 16.26 | 34.26 | 10.87 | 8.07 | 14.46 |
| Unadjusted (Marder et al., 2015) | 60 | 7.84 | 4.61 | 12.27 | 2.79 | 1.80 | 3.99 |
|  | 65 | 12.28 | 7.59 | 17.89 | 4.44 | 3.11 | 6.08 |
|  | 70 | 17.89 | 12.05 | 24.83 | 6.60 | 4.70 | 8.95 |
|  | 75 | 22.16 | 15.03 | 30.39 | 8.31 | 6.13 | 10.99 |
|  | 80 | 26.22 | 17.94 | 36.30 | 10.00 | 7.31 | 13.37 |

The penetrance estimates adjusted for relative's sex and carrier status interaction, proband's sex, and site of enrollment (Proposed) was compared to the penetrance estimates unadjusted for any covariates reported in the previous study (Marder et al., 2015). Denote the estimated age-specific risk (Risk) and 95% confidence intervals (Lower limit and Upper limit).

Table S4: BIC for analyses assuming time-invariant genotype effect $\beta$ (Cox PH model) and time-varying genotype effect $\beta(t)$ estimated by B-splines with various number of knots and degrees in the Ashkenazi Jewish *LRRK2* Consortium study.

| Analysis* | Cox PH model | Time-varying genotype effect | | | | |
| --- | --- | --- | --- | --- | --- | --- |
|  |  |  | Number of knots | | | |
|  |  | degree | 0 | 1 | 2 | 3 |
| Scenario 1 | 91.6 | Linear | 105.9 | 111.5 | 119.6 | 125.3 |
|  |  | Quadratic | 108.8 | 120.4 | 126.8 | 135.8 |
|  |  | Cubic | 120.4 | 129.0 | 137.6 | 144.3 |
| Scenario 2 | 97.0 | Linear | 111.6 | 116.8 | 125.1 | 130.5 |
|  |  | Quadratic | 114.0 | 126.0 | 127.6 | 139.8 |
|  |  | Cubic | 126.0 | 134.8 | 143.3 | 151.5 |
| Scenario 3 | 114.6 | Linear | 129.4 | 134.7 | 143.0 | 148.3 |
|  |  | Quadratic | 131.9 | 143.9 | 145.5 | 157.5 |
|  |  | Cubic | 143.8 | 152.7 | 161.2 | 169.2 |

*: Scenario 1 adjusted for relative's sex and carrier status interaction. Scenario 2 adjusted for relative's sex and carrier status interaction and proband's sex. Scenario 3 adjusted for relative's sex and carrier status interaction, proband's sex, and site of enrollment.

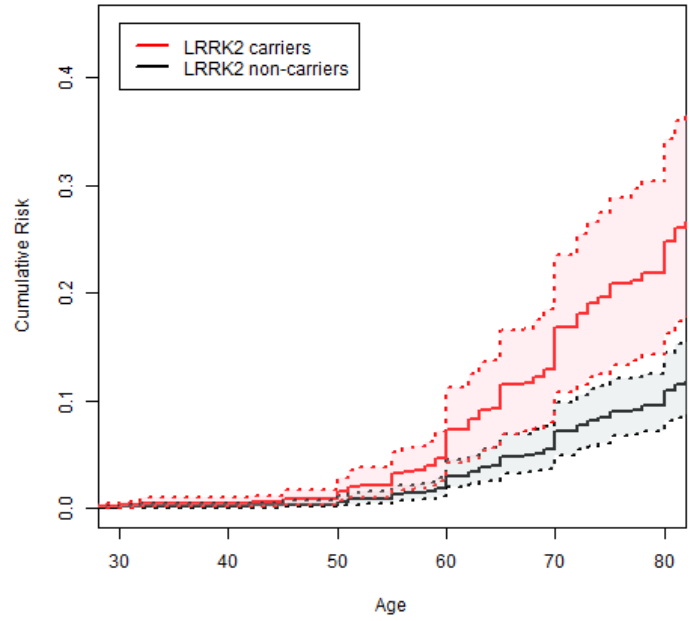

Figure S5: Estimated marginal age-specific risk of Parkinson's disease (PD) in *LRRK2* G2019S carriers (red solid line) and non-carriers (black solid line) with their 95% confidence intervals (dashed lines).